\documentclass{iauc} 
\usepackage{graphicx}

\title[Evolution of Atmospheric Biomarkers] 
{Atmospheric biomarkers and their evolution over geological timescales}

\author[Kaltenegger, Jucks, Traub]   
{Kaltenegger L.$^1$%
  \thanks{Present address:Harvard-Smithsonian Center for Astrophysics, MS20, 60   Garden Street, Cambridge, MA 02138, USA },
Jucks K.$^1$ \and Traub W.$^1$$^,$$^2$}

\affiliation{$^1$Harvard-Smithsonian Center for Astrophysics,
60 Garden St, Cambridge, MA 02138, USA \break 
email: lkaltenegger@cfa.harvard.edu; kjucks@cfa.harvard.edu\\[\affilskip]
$^2$ Jet Propulsion Lab,  M/S 301-451, 4800 Oak Grove Dr., Pasadena CA, 91109, USA \break 
email: wtraub@jpl.nasa.gov}

\pubyear{2005}
\volume{IAUC200} 
\pagerange{1--6}
\date{Nov 30th}
\jname{Direct Imaging of Exoplanets: Science and Techniques}
\editors{C.Aime , F. Vakili, eds.}
\begin{document}

\maketitle

\begin{abstract}
The search for life on extrasolar planets is based on the assumption that one can screen extrasolar planets for 
habitability spectroscopically. The first space born instruments able to detect as well as characterize extrasolar 
planets, Darwin and terrestrial planet finder (TPF-I and TPF-C) are scheduled to launch before the end of the next decade. 
The composition of the planetary surface, atmosphere, and its temperature-pressure profile influence a detectable 
spectroscopic signal considerably. For future space-based missions it will be crucial to know this influence to interpret 
the observed signals and detect signatures of life in remotely observed atmospheres. We give an overview of biomarkers in 
the visible and IR range, corresponding to the TPF-C and TPF-I/DARWIN concepts, respectively. We also give an overview of 
the evolution of biomarkers over time and its implication for the search for life on extrasolar Earth-like planets. We show 
that atmospheric features on Earth can provide clues of biological activities for at least 2 billion years. 


\end{abstract}

\firstsection 
\section{Introduction}
Indirect methods to detect extrasolar terrestrial planets such as radial velocity, astrometry, and transits only allow getting 
morphological information. The planet 
is deduced from its effect on the parent star under survey. To understand the properties of a planet we need 
to collect photons from the planet directly to determine its physical and chemical properties. Spectral analysis of extrasolar planets is a very young science 
that only recently achieved first results for hot Jupiters. Some atmospheric components like Na were detected in absorption in 
the upper atmosphere of HD209458b (\cite[Charbonneau, Brown, Noyes \etal\ (2002)]{Charbonneau02}). The Spitzer telescope recently 
detected the infrared emission from two transiting hot Jupiters by subtracting the stellar flux when the transiting planet 
is eclipsed by its star from the combined flux of the star and planet (see \cite[Deming, Seager, Richardson \etal\ (2005)]{Deming05},  \cite[Charbonneau, Allen, Megeath \etal\ (2005)]{Charbonneau05}). 

The first space born instruments able to detect as well as characterize extrasolar planets are DARWIN (see e.g.  \cite[Kaltenegger, Fridlund \& Karlsson (2005)]{Kaltenegger05proc}), 
TPF-I and TPF-C (see e.g. \cite[Borde \& Traub (2005)]{Borde05}). DARWIN/TPF-I are both based on 
the concept of a free flyer IR interferometer architecture, and TPF-C is envisioned as a coronagraph in the visible. 
The strategy to search for biological activity on terrestrial planets is based on the assumption that one can use 
spectroscopy to screen extrasolar planets for habitability. Basing the search for life on the carbon chemistry assumption 
allows establishing criteria for habitable planets in terms of their size and distance from their stars ( \cite[Kasting, Whitmire \& Reynolds (1993)]{Kasting93}).
Note that observations over interstellar distances can only distinguish those planets whose habitability and biological 
activity is apparent from observations of the reflected or emitted radiation. A crucial factor in interpreting planetary spectra is the point in the evolution of the atmosphere and its 
biomarkers over time. Concentrating on the evolution of our planet we establish a model for its 
atmosphere and the detectable biomarkers over its evolution history. 
Figure~\ref{fig2} shows that atmospheric features on Earth can provide clues of possible life forms for at least 2 billion years (see \cite[Kaltenegger, Traub \& Jucks  (2005)]{Kaltenegger05}).

\subsection{Atmosphere evolution on Earth}
The Earth formed about 4.5 billion years ago. The primitive atmosphere was formed by the release of volatiles from the interior,
and/or volatiles delivered during the late bombardment period. This atmosphere was most likely dominated by carbon dioxide, 
with nitrogen being the second most abundant gas and trace amounts of methane, ammonia, sulphur dioxide, hydrochloric acid 
and oxygen (see \cite[Kasting \& Siefert (2002)]{Kasting02}). Carbon dioxide and/or methane played a crucial role in the development of an early greenhouse 
effect that counteracted the lower solar output. Two major processes have changed the primitive atmosphere, the reduction of 
$CO_2$ and the increase of $O_2$. A huge amount of $CO_2$ must have been removed from the atmosphere, most likely by 
the burial of carbon into carbonate rocks, though the process is still debated. Primitive cyanobacteria are believed to have 
produced oxygen. After most of the reduced minerals were oxidized, about two billion years ago, atmospheric oxygen could 
accumulate. Different schemes have been suggested to quantify the rise of oxygen and the evolution 
of life by anchoring the points in time to fossil finds (see e.g. \cite[Owen (1980)]{Owen80} \cite[Schopf (1993)]{Schopf93} \cite[Ehrenfreund \& Charnley (2003)]{Ehrenfreund03}). There are still many open questions.  We use a climate model based on work by \cite[Kasting \& Catling (2003)]{Kasting03}, \cite[Pavlov, Hurtgen, Kasting, Arthur (2003)]{Pavlov03}, 
\cite[Segura, Krelove, Kasting \etal\ (2003)]{Segura03} and \cite[Traub \& Jucks (2002)]{Traub02} to create a schematic atmospheric model of our Earth over geological timescales. 
We find a surface temperature above the freezing point for all of Earth's history (not considering short term atmospheric 
changes during glaciation periods). Our radiative transfer model is based on the model by \cite[Traub \& Jucks (2002)]{Traub02} that was developed to successfully analyse data from 
stratospheric balloon flight campaigns ongoing during the last few years. The same radiative transfer code was used to 
successfully analyse data from Earthshine measurements in the visible (400 to 900 nm) (see \cite[Woolf, Smith, Traub \etal\ ]{Woolf02}) and mid-infrared (700 to 2400 nm) (see \cite[Turnbull, Traub, Jucks \etal\ (2005)]{Turnbull05}). It shows excellent agreement with measurements in both the visible and IR wavelength band. We present the spectra of Earth
over geological timescale in anticipation of proposed space based terrestrial planet search missions to operate between 
500 to 1100 nm in the visible to mid-IR range and 6 to 20$\mu$m in the IR. Our model shows a considerable 
difference in biosignatures that can be detected with those missions. That should be able to constrain an Earth-like planet 
to its time in evolution.

\subsection{Atmospheres and Biomarkers}
The spectral characteristics of a planet are a very important guide to identifying the best wavelength region to probe 
for a planet as well as characterize the spectra once it has been detected. The TPF Science Working Group (TPF-SWG) 
identified the waveband between 8.5$\mu$m to 20$\mu$m and preferably 7$\mu$m to 25$\mu$m, respectively, for the search for biomarkers in 
the mid-IR region and 0.7$\mu$m to 1.0$\mu$m and preferably 0.5$\mu$m to 1.1$\mu$m for the visible to near IR region (see \cite[Des Marais, Harwit, Jucks, \etal\ (2002)]{DesMarais02}). 
The DARWIN project concentrates on the waveband between 6$\mu$m to 20$\mu$m (see \cite[Fridlund, Kaltenegger, \etal\ (2006)]{Fridlund06}). In the thermal part of the spectrum, 
the shape gives a measure of the temperature of the object examined. Observations from 8$\mu$m to 12$\mu$m of the $H_2O$ continuum allow 
estimations of the surface temperature of Earth-like planets. For higher concentrations of water vapor in the atmosphere, the continuum gets lowered due to water absorption. 

Biomarkers are features whose presence or abundance requires a biological origin (see 
\cite[Des Marais, Harwit, Jucks, \etal\ (2002)]{DesMarais02},  \cite[Meadows \etal\ (2005)]{Meadows05}). 
They are created either during the 
acquisition of energy and/or the chemical ingredients necessary for biosynthesis (e.g. atmospheric oxygen and methane) or 
are products of the biosynthesis (e.g. complex organic molecules and cells). As signs 
of life in themselves $H_2O$ and $CO_2$ are secondary in importance because although they are raw materials for life, they are 
not unambiguous indicators of its presence. Taken together with molecular oxygen, abundant $CH_4$ can indicate biological 
processes. Depending on the degree of oxidation of a planet's crust and upper 
mantel non-biological mechanisms can also produce large amounts of $CH_4$ under certain circumstances.
Oxygen is a chemically reactive gas. Reduced gases and oxygen have to be produced concurrently to be detectable in the 
atmosphere, as they react rapidly with each other. The 9.6$\mu$m $O_3$ band is highly saturated and is thus a poor 
quantitative indicator, but an excellent qualitative indicator for the existence of even traces of $O_2$. Ozone is a very 
nonlinear indicator of $O_2$ because the ozone column depth remains nearly constant as $O_2$ increases from 0.01 present atmosphere 
level (PAL) to 1 PAL. 

$N_2O$ is a very interesting molecule because it is produced in abundance by life but only in trace amounts by natural processes. 
There are no $N_2O$ features in the visible and two weak $N_2O$ features in the IR. In the IR it can only be detected in regions 
strongly overlapped by $CH_4$, $CO_2$ and $H_2O$, so it is unlikely to become a prime target for the first generation of space-based 
missions searching for extrasolar planets that will work with low resolution, but it is an excellent target for follow up 
missions. Spectral features of $N_2O$ become apparent in atmospheres with less $H_2O$ vapor. In the mid-IR $N_2O$ has a band at 7.9$\mu$m, comparable in strength to the adjacent $CH_4$ band but weak compared 
with the overlapping $H_2O$ band. The absorption bands of those three species are different. It is not readily separable for low 
resolution spectroscopy for current Earth, but the methane feature is easily detectable for early type Earth planets 
according to our models. 
To detect and study surface properties we can only use wavelengths that penetrate to the planetary surface. On a cloud-free 
Earth, the diurnal flux variation caused by different surface features rotating in and out of view could be high. 
When the planet is only partially illuminated, more concentrated signal from surface features could be detected as they 
rotate in and out of view. Most surface features like ice or sand show very small or very smooth continuous opacity changes 
with wavelength. Figure~\ref{fig1} shows that the signal detected from a cloud free planet is lower than that from an Earth-like planet. 
Thus the integration time needed to detect such planets will be longer.

\subsection{The red edge of land planets}
An interesting example for surface biomarkers on Earth is the red edge signature from photosynthetic plants at about 750 nm. 
Photosynthetic plants have developed strong infrared reflection as a cooling mechanism to prevent overheating and chlorophyll 
degradation. The primary molecules that absorb the energy and convert it to drive photosynthesis ($H_2O$ and $CO_2$ into sugars 
and $O_2$) are chlorophyll A (0.450$\mu$m) and B (0.680$\mu$m). 
Several groups have measured the integrated Earth spectrum via the technique of Earthshine, sunlight reflected from 
the ``dark'' side of the moon. Earthshine measurements have shown that detection of Earth's vegetation-red edge (VRE) is feasible 
but made difficult owing to its broad, essentially featureless spectrum and cloud coverage. Trying to identify such weak, 
continuum-like features at unknown wavelengths in an extrasolar planet spectrum requires models for different planetary 
conditions. Our knowledge of different surface reflectivities on Earth, like deserts, ocean and ice, help assigning the 
VRE of the Earthshine spectrum to terrestrial vegetation (see Figure~\ref{fig2}). Earth's hemispherically integrated vegetation red-edge signature is weak, but planets with different rotation 
rates, obliquities, land-ocean fraction, and continental arrangement may have lower cloud-cover and higher vegetated fraction (see e.g. \cite[Seager (2002)]{Seager02}). 

\subsection{Model results}

\begin{figure}
\centering
\resizebox{4.6cm}{!}{\includegraphics{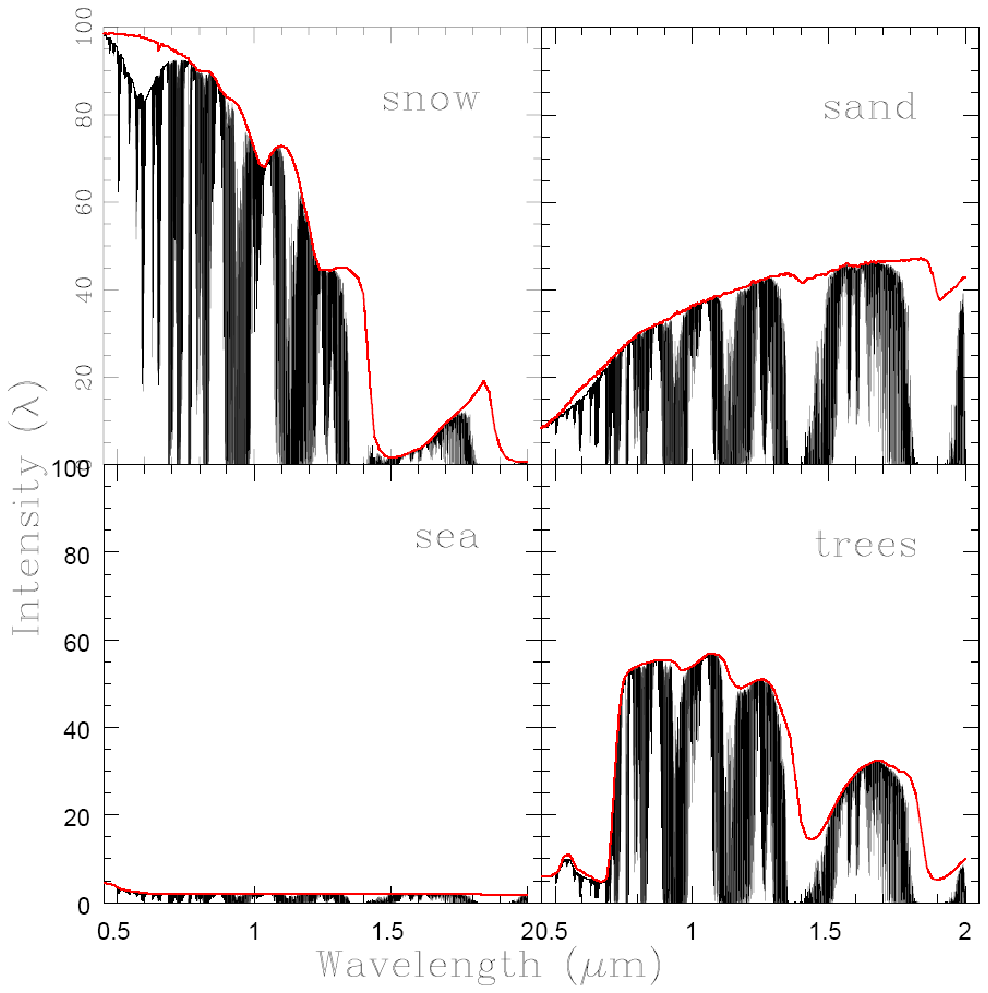} }
\resizebox{4.75cm}{!}{\includegraphics{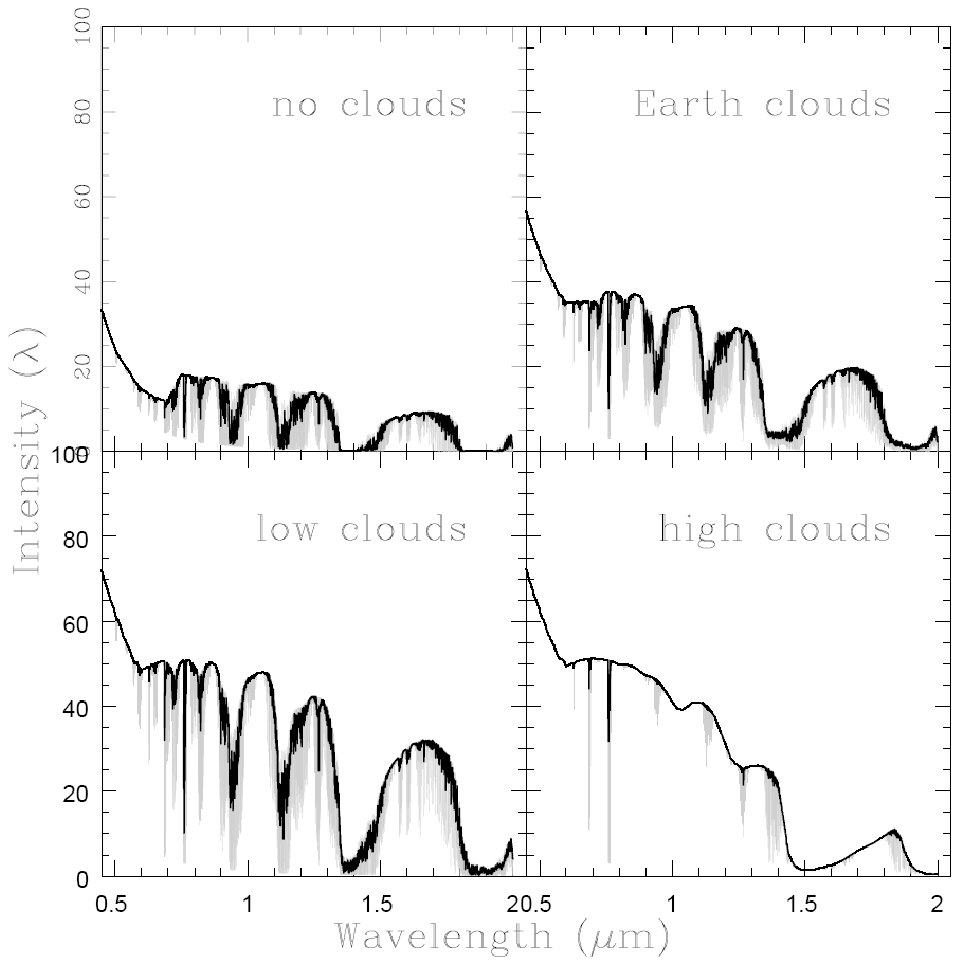} }
  \caption{(left) Signal of a current Earth atmosphere considering surface features assuming a clear atmosphere without clouds; Different surface composition (a) snow (b) sea (c) trees (d) sand. (right) Signal of current Earth with realistic surface features and (a) no clouds, (b) 100\% cumulus cloud coverage at 4 km and (c) 100\% cirrus cloud coverage at 12 km and (d) a mixture of clouds resembling the present Earth. }\label{fig1}
\end{figure}

\begin{figure}
\centering
\resizebox{5.7cm}{!}{\includegraphics{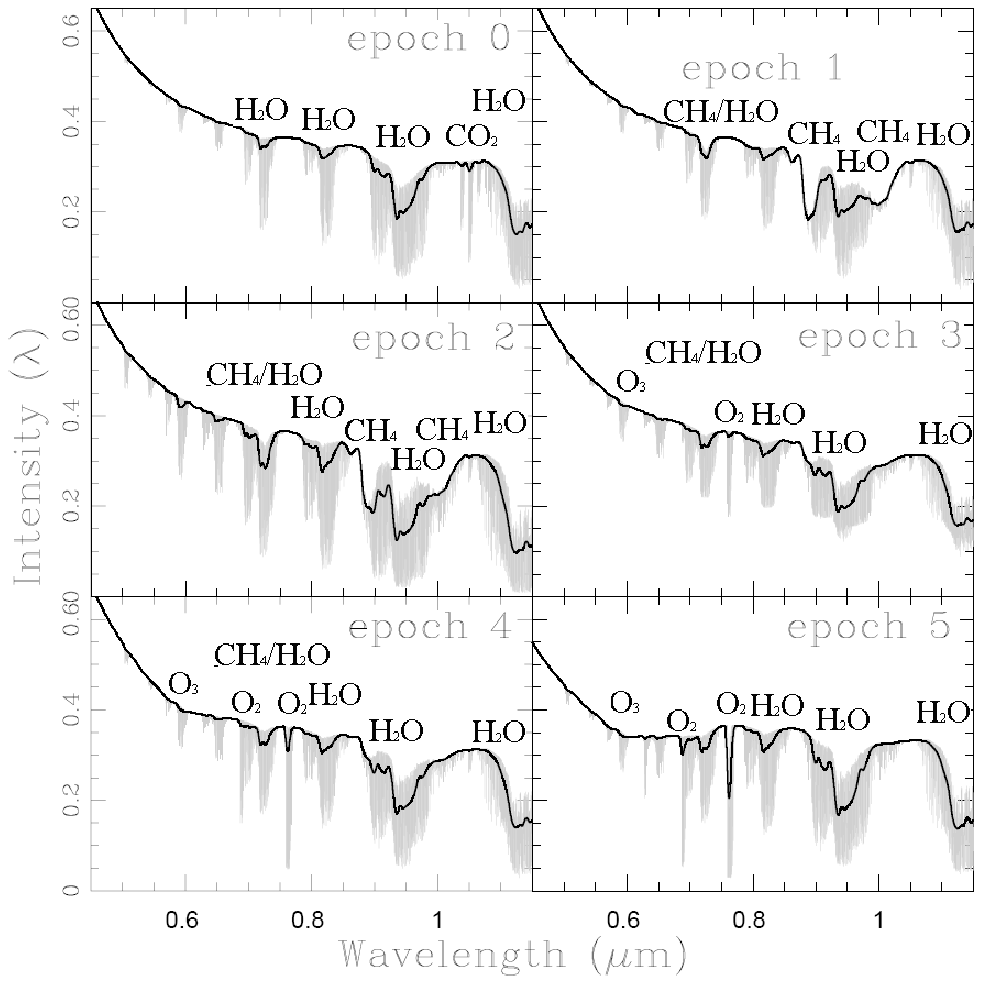} }
\resizebox{5.7cm}{!}{\includegraphics{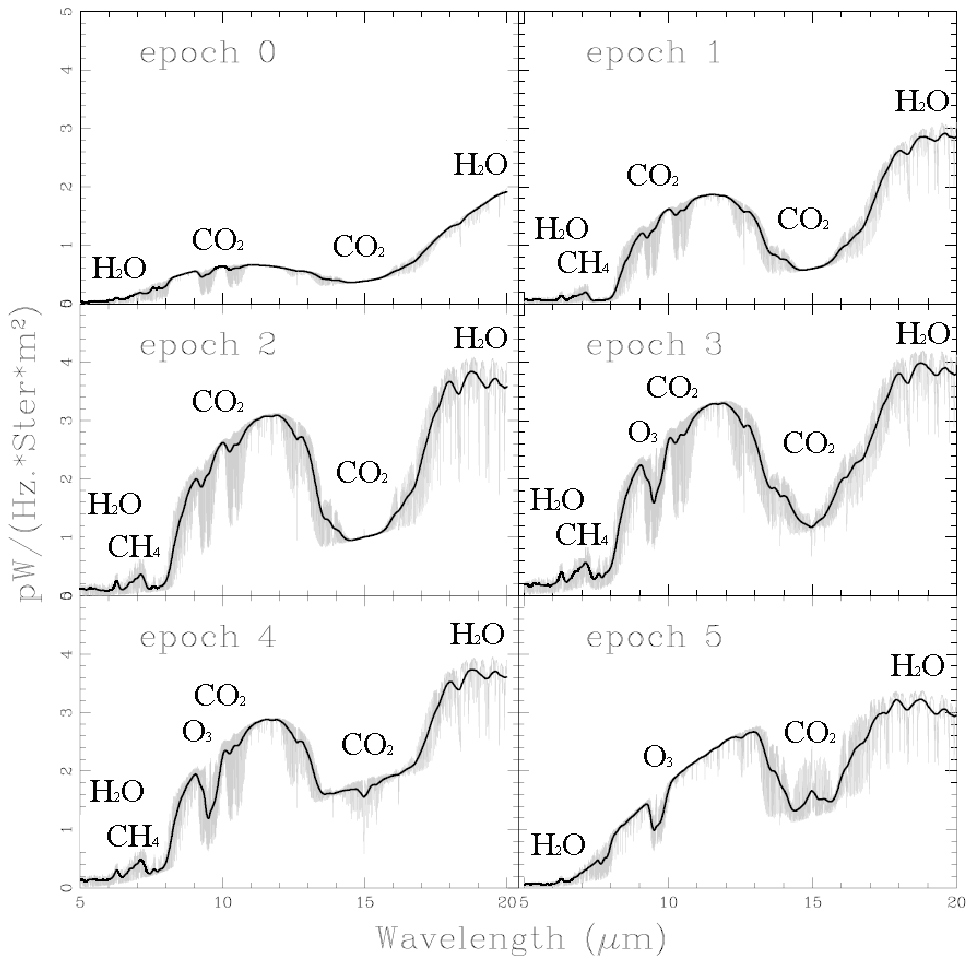} }
   \caption{The atmospheric features on an Earth-like planet change considerably over its evolution from a current day atmosphere (epoch 5 = present Earth) over a $CO_2$/$CH_4$-rich atmosphere (epoch 3) to a $CO_2$ rich (3.9Ga ago = epoch 0) in both the visible (left) and the IR range (right). The black lines show spectral resolution of 70 for the visible and 25 for the IR, comparable to the proposed mission concepts. }\label{fig2}
\end{figure}

Our calculations (see \cite[Kaltenegger, Traub \& Jucks  (2005)]{Kaltenegger05}) show that the composition of the surface (especially in the visible), the atmospheric composition and 
temperature-pressure profile can all have a significant influence on the detectability of a signal. 
The reflectivity of the Earth has not been static throughout the past 4.5Ga (Ga = $10^9$ years ago). Oxygen and ozone became 
abundant about roughly 2.3Ga, affecting the atmospheric absorption component of the reflection spectrum. About 2Ga, a green 
phytoplankton signal developed in the oceans and about 0.44Ga, an extensive land plant cover developed, generating the red 
chlorophyll edge in the reflection spectrum. The oxygen and ozone absorption features could have been used to derive the 
presence of biological activity on Earth anytime during the past 50\% of the age of the solar system, while the chlorophyll 
red-edge reflection feature evolved during the most recent 10\% of the age of the solar system. Figure~\ref{fig1} shows the spectral signature of an Earth atmosphere assuming different surface composition and the big impact clouds can have on the 
spectrum. From the 4 surfaces 
plotted on the left one sees that an Earth-size ocean planet will be more difficult to detect than a snow covered planet 
due to its low albedo. The smooth lines in the left panel of Figure~\ref{fig1} show the albedo of the surfaces without overlayed 
atmosphere. The panel on the right in Figure~\ref{fig1} shows the strong impact clouds can have on the detected signal. Present Earth 
has about 60\% cloud coverage. 100\% cumulus (low) and 100\% cirrus (high) cloud coverage, respectively, are shown for reference. A planet without clouds or snow has a lower overall albedo and thus 
requires a longer integration time for detection than a planet with clouds.  
Figure~\ref{fig2} shows Earth's reflected and thermal emission spectra respectively with its major molecular species ($H_2O$, $O_3$, $O_2$, 
$CH_4$, $CO_2$, $N_2O$) as well as minor contributors ($H_2S$, S$O_2$, $NH_3$, $SF_6$, CFC-11, CFC-12).
The dark lines show a resolution of 70 in the visible and 25 in the IR, as proposed for the TPF-C and Darwin/TPF-I mission, 
respectively. The atmospheric features on an Earth-like planet change considerably over its evolution from a current day 
atmosphere (epoch5 = present Earth) over a $CO_2$/$CH_4$-rich (epoch 3 around 2Ga) to a $CO_2$ rich (3.9Ga = epoch 0) atmosphere. $O_3$ shows a strong feature in the IR for the last 2Ga of Earth's history even at a low resolution of 25. $O_2$ shows a 
weaker feature in the visible for the last 2Ga of Earth's history at a resolution of 70 (note that no noise level is 
assumed for these calculations). 

In the mid-IR the classical signatures of biological activity are the 9.6$\mu$m $O_3$ band, the 15$\mu$m $CO_2$ band and the 6.3$\mu$m $H_2O$ 
band or its rotational band that extends from 12$\mu$m out into the microwave region (\cite[Selsis, Despois, Parisot (2002)]{Selsis02}). For the visible, a similiar investigation has not been conducted yet. In the same spectral region, the 7.7$\mu$m 
band of $CH_4$ is a potential biomarker for early-Earth type planets. In the visible to near-IR one can see a strong $O_2$ absorption feature at 0.76$\mu$m, a broadband $O_3$ absorption at 0.45$\mu$m to 
0.75$\mu$m and a strong $H_2O$ band at 0.94$\mu$m. The strongest $O_2$ feature is the saturated Fraunhofer A-band at 0.76$\mu$m that is still 
relatively strong for significantly smaller mixing ratios than present Earth's. $O_3$ has two broad features, the extremely 
strong Huggins band in the UV shortward of 0.33$\mu$m and the Chappius band which shows as a broad triangular dip in the middle of the 
visible spectrum from about 0.45$\mu$m to 0.74$\mu$m. Methane at present terrestrial abundance (1.65ppm) has no significant visible 
absorption feature, but at high abundance it has strong visible bands at 0.88$\mu$m and 1.0$\mu$m, readily detectable in early Earth's 
history. $CO_2$ has negligible visible features at present abundance, but in a high $CO_2$-atmosphere of 10\% it would 
have a significant band at 1.2$\mu$m and even stronger ones at longer wavelengths. 
Especially in the early evolution stage, the weak 1.06$\mu$m band can be observed (epoch 0).

%


\section{Conclusions}\label{sec:concl}
Concentrating on the evolution of our planet we established a model for its atmosphere and the detectable biomarkers over 
its evolution history and showed the results for TPF-I/Darwin and TPF-C. These missions should be able to constrain an 
Earth-like planet in its evolution. Our calculations show that the atmospheric and the surface composition (especially in 
the visible) and cloud coverage influence the detected signal considerably (see Figure~\ref{fig1}).  From the 4 surfaces plotted on the left 
one sees that e.g. an Earth-size ocean planet will be more difficult to detect than a snow covered planet due to its low albedo (here we consider no clouds). The atmospheric features on an Earth-like planet change considerably over its evolution from 
a current day atmosphere (epoch 5 = present Earth) over a $CO_2$/$CH_4$-rich atmosphere (epoch 3 around 2Ga) to a $CO_2$ rich 
(3.9Ga = epoch 0) see Figure~\ref{fig2}. Atmospheric features on Earth can provide clues of biological activities for at least 
2 billion years. $O_3$ shows a strong feature in the IR for the last 2Ga of Earth's history even at a low resolution of 25. 
$O_2$ shows a weaker feature in the visible for the last 2Ga of Earth's history at a resolution of 70 (note that no noise level 
is assumed for these calculations). Methane (a potential bioindicator for early Earth) and nitrous oxide (a biomarker with
a weak signature in the IR) have features nearly overlapping in the 7$\mu$m region, lying in the red wing of the 6$\mu$m water band. 
The absorption bands of those three species are different. It is not readily separable for current Earth at low spectral 
resolution. $N_2O$ is a very interesting molecule because it is produced in abundance by life but only in trace amounts by natural processes. It is unlikely to become a prime target for the first generation of space-based 
missions searching for extrasolar planets that will work with low resolution, but it is an excellent target for follow up 
missions.
But if methane is abundant, as it probably was for early Earth, then it is readily detectable, even at low 
spectral resolution, functioning as an early bio-indicator for young Earth-like planets.

\vspace{-5pt}
\begin{acknowledgments}
Special Thanks to J. Kasting, A. Knoll and A. Segura for constructive discussions.
\end{acknowledgments}

\end{document}